\documentclass[a4paper,11pt]{article}
\pdfoutput=1 

\usepackage{jinstpub} 
\usepackage{subfigure}

\title{\boldmath Dose rate effects in the radiation damage of the plastic scintillators of the CMS Hadron Endcap calorimeter}

\author[a]{\small V.~Khachatryan} 
\author[a]{\small A.M.~Sirunyan} %
\author[a]{\small A.~Tumasyan}%
\affiliation[a]{\small Yerevan~Physics~Institute,~Yerevan,~Armenia} %

\author[b]{\small A.~Litomin}%
\author[b]{\small V.~Mossolov}%
\author[b,1]{\small N.~Shumeiko\note{deceased}}
\affiliation[b]{\small National~Centre~for~Particle~and~High~Energy~Physics,~Minsk,~Belarus}%

\author[c]{\small M.~Van~De~Klundert}
\author[c]{\small H.~Van~Haevermaet}
\author[c]{\small P.~Van~Mechelen}
\author[c]{\small A.~Van~Spilbeeck}
\affiliation[c]{\small Universiteit~Antwerpen,~Antwerpen,~Belgium}

% ...

\emailAdd{eno@umd.edu}

\abstract{
We present measurements of the reduction of light output by plastic scintillators irradiated in the CMS detector
during the 8 TeV run of the Large Hadron Collider 
and show that they indicate a strong dose rate effect.  
The damage for a given dose is larger for lower dose rate exposures.
The results agree  with previous measurements of dose rate effects,
but are stronger due to the very low dose rates probed.
We show that the scaling with  dose rate is consistent with that expected from diffusion effects.
}


\arxivnumber{1608.07267} 

\collaboration[c]{on behalf of CMS-HCAL collaboration}

\begin{document}
\maketitle
\flushbottom

\section{Introduction}

Plastic scintillators are widely used in detectors for experiments in high energy physics experiments
due to their high light output, low cost, and versatility.  However, they are also known to suffer from radiation damage
(for a detailed review, see~\cite{sauli}). 
 Typically, the light output of the scintillator
decreases exponentially with the dose received, as in Eq.~\ref{eqn:exp}:

\begin{equation}
L(d)  = L_0 \exp(-d/D)
\label{eqn:exp}
\end{equation}

where $L(d)$ is the light output after receiving a dose $d$, $L_0$ is the light output before irradiation, and $D$ is the exponential constant.  The exponential constant $D$ depends on the materials used in the 
construction of the scintillator and on its environmental history.  Several results have also
shown a dependence on dose rate~\cite{sauli,Biagtan1996125,34504,Wick1991472,289295,173180,173178,Giokaris1993315}.
The lowest dose rates probed by these measurements were a few krad/hr.

In this paper, we present results from scintillators used in the CMS hadron endcap calorimeter 
(HE)~\cite{HCALTDR1997}
irradiated at dose rates between $\approx$ 0.1 to a few times $10^{-4}$ krad/hr
that indicate that the radiation damage can have a very strong dose rate dependence.  
We also present a measurement from an irradiation
at a ${\rm ^{60}Co}$ source at a dose rate of 0.28 krad/hr. 

Plastic scintillators consist of a plastic substrate, often polystyrene (PS) or polyvinyltoluene (PVT),
into which wavelength shifting (WLS) fluors have been dissolved, usually
a primary and a secondary fluor. When a charged particle traverses
the scintillator, the molecules of the substrate are excited.  This excitation can be transferred to
the primary fluor radiatively in the deep UV at low concentrations or via the F{\"o}rster mechanism~\cite{forster} at 
higher concentrations.  The primary fluor transfers the excitation radiatively to the secondary
fluor.  De-excitation of the secondary fluor generally produces light in the visible range, to match
well with currently available photodetectors.  The light must traverse the scintillator
to reach the photodetector, and can be absorbed by ``color centers'' along its path.

Dose rates effects have been associated with oxygen diffusion~\cite{seguchi, Wick1991472,Busjan199989}.
The rate of diffusion depends on the substrate material and 
has a weak dependence on dose and on environmental factors~\cite{seguchi}.
For unirradiated plastic, the diffusion rate for 
oxygen is 13 times slower for PVT than for PS~\cite{o2diff}. 
Oxygen can increase the mobility of radicals  created during irradiation when
chemical bonds in the polymer of the substrate are broken,
and can interact with radicals to 
affect the formation of color centers, which absorb light~\cite{Busjan199989}.  
When oxygen is not present, cross linking is enhanced, and gel formation is reduced~\cite{seguchi}.
These processes can affect the energy levels
of the substrate, thus affecting the transfer mechanisms of the initial excitation produced by
the charged particles being measured.
When radicals are more mobile, they are more likely to reform good bonds, reducing their ill effects.
(For a detailed review of radicals in polymers, see \cite{radicals}).
Those color centers that go away after a recovery period, whose length depends on temperature and the 
concentration of oxygen, are referred to as
temporary damage.  Color centers that remain after recovery are referred to as permanent damage.
Oxygen tends to decrease temporary damage but increase permanent damage~\cite{sauli}.

When the concentration of radicals is low, the penetration depth of oxygen into the substrate goes as~\cite{seguchi}

\begin{equation}
z_0^2 = \frac{G}{R}
\end{equation}
where
\begin{equation}
G = \frac{2VSP}{\Phi},
\end{equation}
$R$ is the dose rate,
$V$ is the diffusion constant of the gas, $S$ is the solubility constant of oxygen, $P$ is the oxygen pressure, and
$\Phi$ is the specific rate constant of active site formation.
$G$ is roughly independent of dose~\cite{seguchi,Wick1991472}.

For a simple configuration
consisting of a piece of plastic scintillator with an alpha source on one side and a 
photodetector on the other,
assuming the fraction of the thickness of the scintillator  penetrated by
 the alpha is small, the light produced will traverse first a region
penetrated by oxygen of depth $z_0$, then a region the oxygen does not reach, of thickness ${t-2z_0}$
where $t$ is the thickness of the piece of scintillator, and finally another region with oxygen, before
reaching the photodetector.  If the inverse of the light absorption length
when color centers are formed  in the presence of oxygen is
$\mu_1$ and that formed independent of the presence of oxygen is $\mu_2$, the light output will be

\begin{equation}
L  = L_0 \exp\left( -\mu_1 (2z_0) - \mu_2 t \right).
\end{equation}

The inverse absorption length $\mu$ is related to the density of color centers $Y$
and the cross section for absorption of the light by the color center $\sigma$ by

\begin{equation}
\mu = Y \sigma.
\end{equation}

When the color center density $Y$ is low, 
it depends on the chemical yield $g$, the density of the scintillator $Q$, and the
dose $d$ as

\begin{equation}
Y=gQd.
\end{equation} 

The light output is then

\begin{equation}
L(d) = L_0 \exp\left( - \left(g_1 Q_1 \sigma_1 2 \frac{\sqrt{G}}{\sqrt{R}}\right)d - \left(g_2 Q_2 \sigma_2 t \right)d  \right)
\end{equation}

and so the functional form of the dependence of the exponential constant on the dose rate is

\begin{equation}
D = \frac{\sqrt{R}}{A+B\sqrt{R}}
\end{equation}
where $A=2 g_1 Q_1 \sigma_1 \sqrt{G}  $ and $B=g_2 Q_2 \sigma_2 t$.

For a more complex arrangement, the dependence will not be as simple, but still might be expected to
show a $\sqrt{R}$ dependence at low dose rate and to approach a constant value at high dose rate.

\section{Results from the CMS HE laser calibration}

\begin{figure}[hbt]
  \begin{center}
    \includegraphics[width=0.95\textwidth]{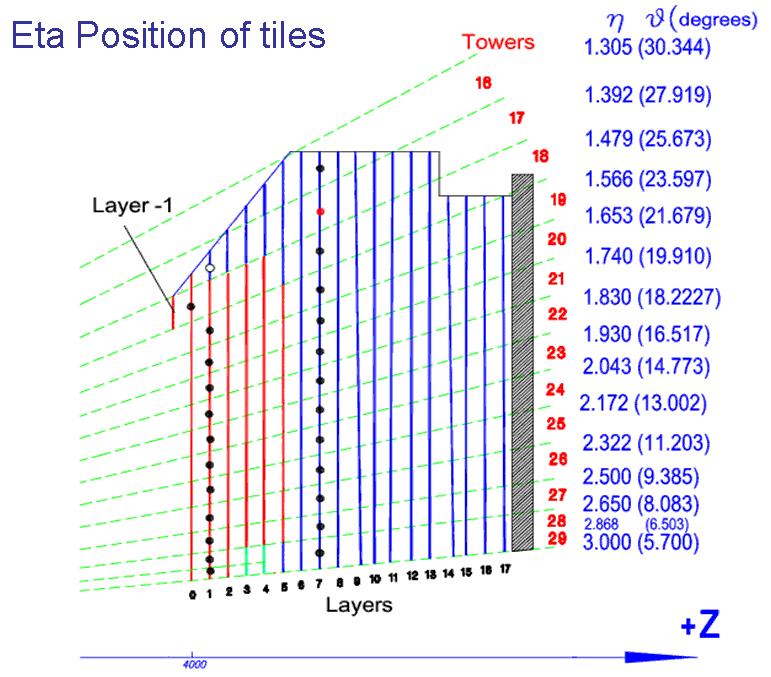}
    \caption{Schematic of the CMS hadron endcap calorimeter.
    }
    \label{fig:he}
  \end{center}
\end{figure}

The HE is part of the CMS detector~\cite{Chatrchyan:2008zzk} at CERN's Large Hadron Collider (LHC).
It is a sampling calorimeter that 
uses brass as its passive material and scintillating tiles as its active material.
It has 18 layers of active material, denoted layers 0 through 17,
over most of its rapidity coverage.
The zeroth layer of scintillator uses BC-408, a PVT-based scintillator from the Bicron division
of the Saint Gobain corporation\footnote{Saint-Gobain Crystals, Courbevoie, France.},
while the other layers use SCSN-81, a PS-based scintillator from Kuraray\footnote{Kuraray Corporation, Otemachi, Chiyoda-ku, Tokyo 100-8115, Japan.}.
Figure~\ref{fig:he} shows a schematic of the HE calorimeter.

\begin{figure}[hbt]
  \begin{center}
    \includegraphics[width=0.95\textwidth]{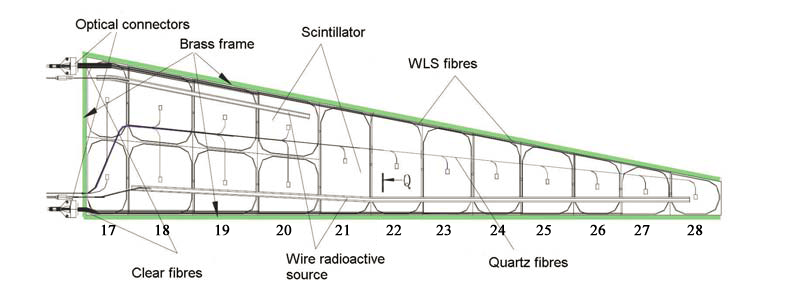}
    \caption{Schematic of the CMS hadron endcap calorimeter scintillator megatiles tiles for tiles in layer 11 through 13.  
The numbers below the tiles represent the tower number.
The megatiles in layers 0 through 4 have an additional tower at large $\eta$ (tower 29).
The megatiles in layers 5 through 10 contain tile 16 at low $\eta$.  Layers 14 through 17 do not have tile 17.
    }
    \label{fig:tile}
  \end{center}
\end{figure}

The tiles are trapezoidal in shape. 
They contain a sigma groove that holds a 0.94~mm Y-11 (Kuraray) WLS fiber,
mirrored on one end.  Fig.~\ref{fig:tile} shows a schematic of the tiles,
arranged in a ``mega tile'' that holds individual tiles.
The calorimeter spans the pseudorapidity region from 1.305 to 3.
The tiles are labeled by their pseudorapidity by ``tower number'',
starting with 16 for the ones at lowest pseudorapidity and going
to 29.  Due to the geometry of the detector, layers 0 through 4
contain tiles corresponding to towers 17 through 29.  Layers 5 through 17 do not have 
tower 29.
Only Layers 5 through 10 contain
tower 16, and layers 14 through 17 do not have tile 17 as well.
The megatiles are inserted into the brass absorber of the HE.
The tile thickness is 0.9~cm in layer 0 and 0.37~cm in the rest of the layers.  The
sizes range from roughly ${\rm 10~cm \times 10~cm }$ to about ${\rm 20~cm \times 20cm }$.
Clear fibers carry the light to hybrid photodiodes (HPD)~\cite{Cushman},
and each HPD signal is digitized by a Charge Integrator and Encoder (QIE)~\cite{qie}.

The light output of the tiles was measured during the 2012 8 TeV run of the LHC.
The total delivered integrated luminosity
was 23.3 ${\rm fb^{-1}}$, and was estimated as described in~\cite{lumi}.
The time was estimated three different ways: using LHC 2012 run performancy summaries that included information
   on total delivered luminosity and hours in stable beam,
using detailed CMS Web Based Monitoring records for each fill parameter
   including delivered luminosity and time in fill,
and using
the peak luminosity, luminosity lifetime, and fill length values to correlate
average lumi rate with peak lumi.
All three methods delivered very similar results.
The instantaneous luminosity was fairly constant
during the run.


The light was measured using
a laser calibration system, consisting of a triggerable nitrogen laser, a system
of neutral density filters, and a light distribution system that delivers the UV light
to the scintillator tiles in layers 1 and 7  via quartz fibers.  Light was also injected directly
into the HPDs.  
The laser light was injected during the run at times without collisions, between fills of the
accelerator with protons.

\begin{figure}[hbtp]
  \begin{center}
   \subfigure[Layer 1]{\includegraphics[width=0.47\textwidth]{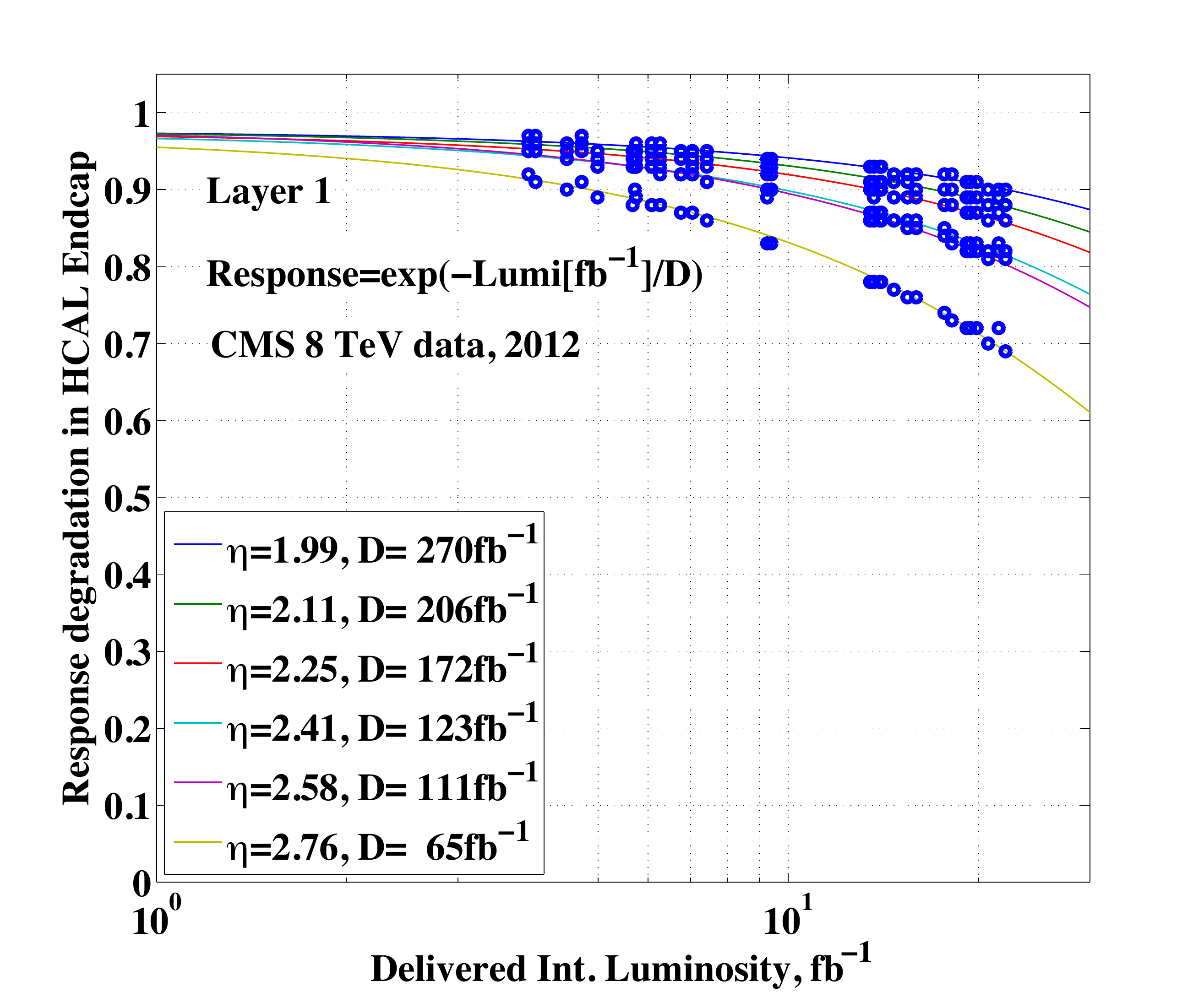}}
   \subfigure[Layer 7]{\includegraphics[width=0.47\textwidth]{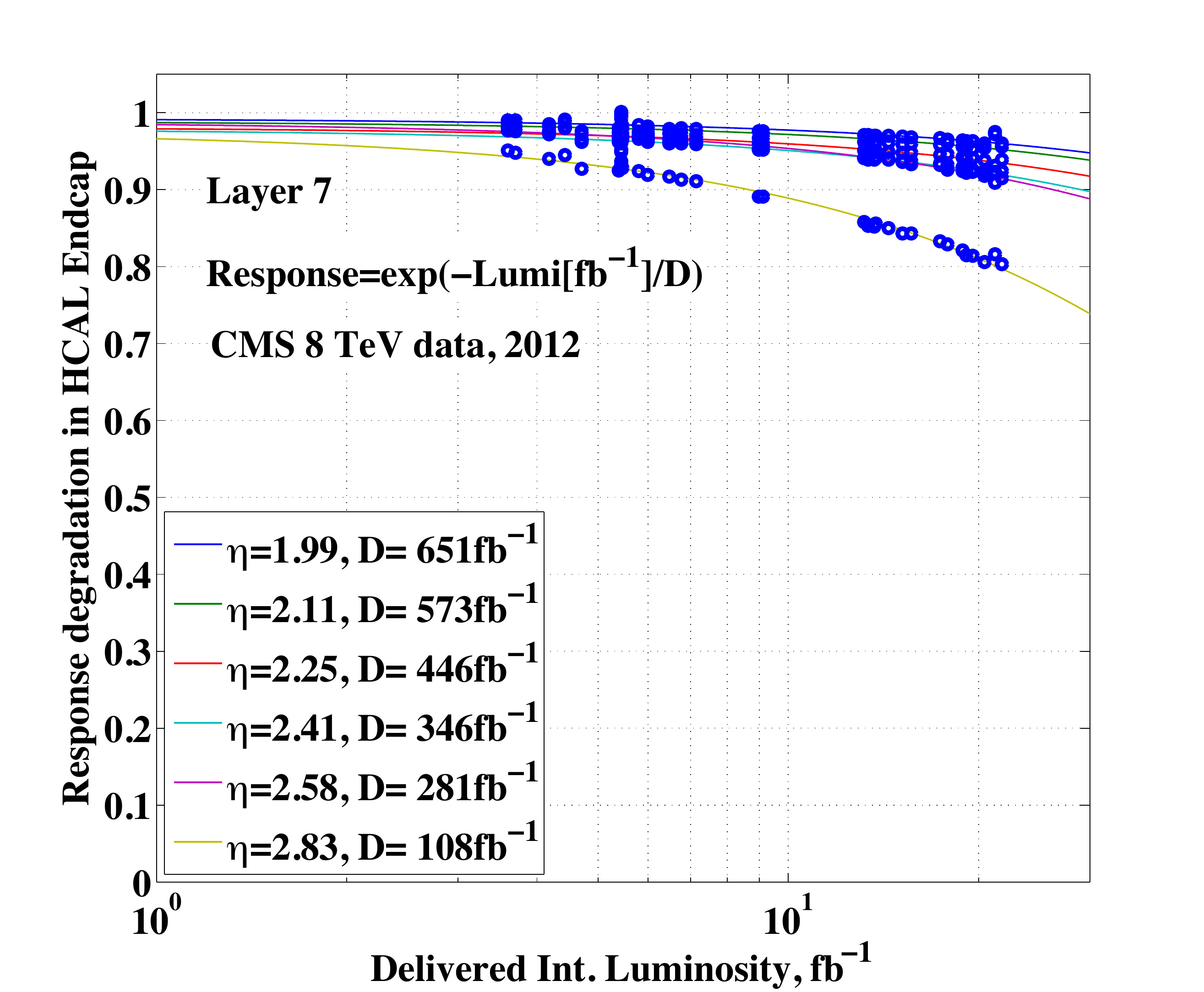}}
    \caption{Ratio of light output to initial light output for tiles as a function of integrated luminosity, with extracted dose constant,
for CMS hadron endcap calorimeter scintillators in Layer 1 (a) and in Layer 7 (b).
    }
    \label{fig:lasercalibL1}
  \end{center}
\end{figure}

Plots of the light output relative to the initial light output
as a function of the accumulated integrated luminosity during the run are shown in 
Fig.~\ref{fig:lasercalibL1}.
They show an exponential decrease in light output
with integrated luminosity.  
After the end of the run, a few percent recovery was observed for the tiles with the largest damage.
The data are not corrected for this effect.

Cross checks of the results from the laser calibration
from calculations of the jet energy scale, a calibration using a ${\rm ^{60}Co}$ source after the end of the run, 
and a measurement looking at the energy distribution in the towers using data taken
with a single electron trigger as a function of integrated luminosity
give similar albeit less precise results.

\begin{sloppypar}
To convert the exponential constant 
in terms of integrated luminosity to an exponential constant
in terms of dose, the dose received by the tile per unit integrated luminosity is needed.
Predictions of the absorbed dose in HE scintillator layers were obtained using the
Monte Carlo code FLUKA 2011.2c~\cite{fluka1,fluka2}.  
The FLUKA predictions for  collisions 
use a model that represents the HE in detail,
with brass, Dural (Aluminium, Cu, Mg, and Mn), Tyvek, 
air, and scintillator layers.  
The absorber regions are represented as 
80~mm thick brass
'LK75' layers (with the exception of the last which is $<$ 20~mm), with a 
density of 8.4 g/${\rm cm^3}$ and a fractional mass composition as follows: Cu 75\%, Zn 24.533\%, 
Si 0.3\%, P 0.01\%, Fe 0.1\%, Sb, 0.005\%, Pb 0.05\%, Bi 0.002\%.  
Layer 0 is a 10~mm thick polyvinyltoluene layer
modelled with a density 1.032 g/${\rm cm^3}$ and fractional mass composition: H 8.5292\% and C 
91.4708\%.  Other scintillator regions (layers 1-17) are 4~mm thick and 
represented with a polystyrene plastic scintillator 
of density 1.05 g/${\rm cm^3}$ and 
fractional mass composition H 7.7423\% and C 92.2577\%.  
Since the energy loss per mass per unit area is more than a factor two higher for hydrogen than for most other materials, the spatial resolution used in the calculation was defined so that the dose estimates were not averaged over regions containing materials other than scintillator layers.
For towers 28 and 29, the amount of material in front of the HE varies with azimuthal angle due,
primarily, to
to the rectangular shape of the crystals in the electromagnetic calorimeter.
This irregularity is not yet simulated, and the dose is calculated for the average
instead.  
The dose was calculated using an R-Phi-Z  mesh, 
independent of the geometry,  overlaying the HE region.  The 
resolution was selected assuming a phi symmetry and aligned in order to obtain dose values 
within the scintillator regions only. For the HE there was one phi bin in total, 
a one mm resolution in Z, and a one cm resolution in R. 
Predictions for the 4-TeV-per-beam '2012' run use a ``Run1'' CMS FLUKA model, 
with a  central beam 
pipe and muon endcap region which is modelled to reflect the configuration at that time.

Since the dose received varies over the surface
of the tile, the rapidity-averaged value is used.  Using a value
averaged over the radius of the tile gives similar results.  
For layer 1, the dose received by a tile ranged from 0.01 to 0.2~Mrad.
For layer 7, the dose received ranged from 0.00005 to 0.03~Mrad.

The doses and dose rates are highest for the tiles closest to the beam line.  
If the exponential constant $D$ does not depend on dose rate,
we would expect the tiles to have the same exponential constant, regardless of dose.
Monte Carlo studies based on the optical transport code in GEANT4~\cite{geant} shows that the exponential constant does not
depend strongly on the tile size.  
Measurements confirm the results of the simulation. Prior to LHC turn on, tiles constructed of SCSN-81 with 
dimensions of ${\rm 5~cm \times 8~cm }$, ${\rm 12~cm \times 8~cm }$, and ${\rm 20~cm \times 20~cm }$ were irradiated at a 
${\rm ^{60}Co}$ source at 
Argonne National Laboratory at a relatively high dose rate of 100 krad/hr. 
The resulting dose constants were 1.9, 1.4, and 0.9 Mrad, respectively. The variation with size is much 
smaller than that seen with the in situ measurements~\cite{pawel}.
The extracted dose constants are not corrected for this effect.
\end{sloppypar}

\begin{figure}[hbt]
  \begin{center}
    \includegraphics[width=0.95\linewidth]{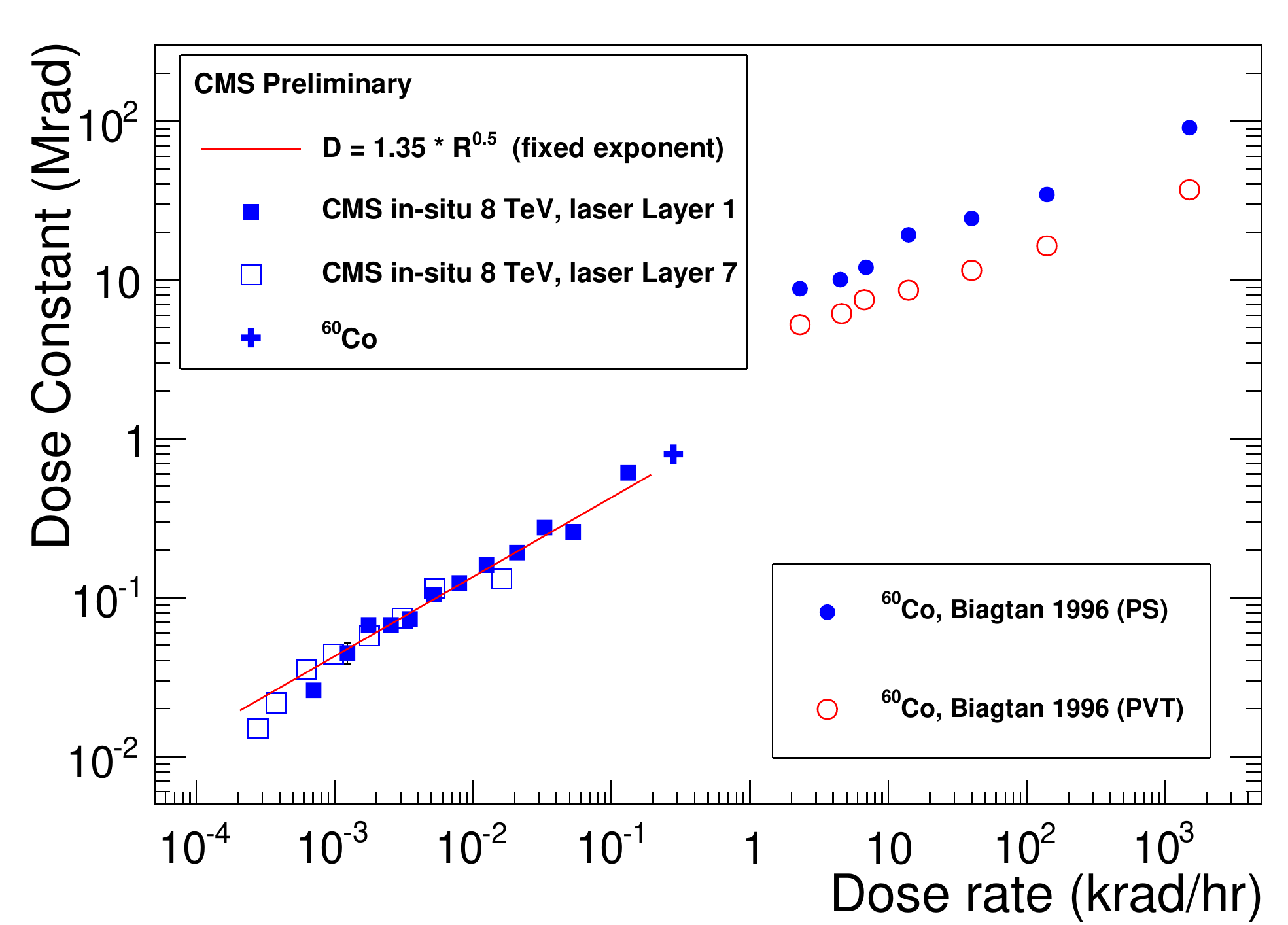}
    \caption{Exponential constant as a function of dose rate.  Results from scintillators based on PS are shown
in blue, while those based on PVT are shown in red.  Results from Layer 1 (7) of the CMS HE scintillator
are indicated by filled squares (open squares).  Each point corresponds to a different tile pseudorapidity.
Results from the ${\rm ^{60}Co}$ irradiations discussed in
Biagtan~\cite{Biagtan1996125} are indicated
with circles.
The label ``${\rm ^{60}Co}$, CMS'' (cross) refers to the measurement first presented in
this paper, which was taken using irradiation from a gamma source.
A fit to the in situ data to a power law is shown, where the power is set
to 0.5.
    }
    \label{fig:DoseConstVsDoseRate}
  \end{center}
\end{figure}

The extracted exponential constant in Mrad is shown in Fig.~\ref{fig:DoseConstVsDoseRate}.
The different points, at different dose rates, correspond to different tiles in layers 1 and 7.
We see a strong dependence on dose rate, with the tiles with the lowest dose rate
having the lowest exponential constant and thus suffering more damage for the same dose than
the tiles with higher exponential constants.

We fit the extracted $D$ values from the laser results
to a power law with an exponent of 0.5.
The results of the fit are included in Fig.~\ref{fig:DoseConstVsDoseRate}.  
The agreement is good.


\section{Results from a low dose rate irradiation at a \texorpdfstring{${\rm ^{60}Co}$}{Co-60} source}

A rectangular tile with WLS fiber,
whose construction was similar to those used in the HE (SCSN-81, and with dimensions of 
10~cm  by 10~cm, with a 50~cm long mirrored fiber)  was irradiated in air
using the Michigan Memorial Phoenix   ${\rm ^{60}Co}$ source at the University of Michigan, 
Ann Arbor, MI.
The integrated dose was 300 $\pm$ 30 krad
accumulated over an irradiation time of 1080 hours,
for a dose rate of 0.28~krad/hr.
The light output when the tile is exposed to a (different) ${\rm ^{60}Co}$ source was measured
before and after irradiation using
a Hamamatsu R580-17 PMT coupled to a Keithley picoammeter.
An unirradiated identical tile was used to check the stability of the measurement system.
The tile was allowed to recover for 6 days
between the exposure and the ``after'' measurement, so as to measure only the permanent damage.

The resulting exponential constant is shown in Fig.~\ref{fig:DoseConstVsDoseRate}.
The dose rate for this measurement was larger than that for any tile in the in situ measurement,
and
the resulting exponential constant is larger than that obtained from the in situ measurements
but much smaller than values obtained from high dose rate reactor exposures discussed in the next section.
Its value agrees well with the extrapolation of the trend from the in situ measurements
to its dose rate.

\section{Comparison to previous results}

We compare our results to those from Ref.~\cite{Biagtan1996125}.  
The authors studied both SCSN-81, which is PS-based, and Bicron-499-35, which is PVT-based.
They irradiated disks with a 0.4~cm thickness and a 1'' diameter with a gamma source.
The light output was measured using an alpha source, after allowing
time for recovery of the temporary damage.  The exponential constant
$D$ was obtained, for dose rates below 14 krad/hr, where measurements were made for only one dose value,
by reading the values of their data from their graphs of light output, and
using Eq.~\ref{eqn:exp} to solve for $D$. For higher dose rates,  where measurements were made for multiple values
of the total dose $d$,
the values of the light output versus dose rate
are obtained from the functional form in their Table 1.  Values with the same dose rate but different dose are then
fit with an exponential to obtain $D$.
%
%

The values from Ref.~\cite{Biagtan1996125} have
exponential dose constants of
tens of Mrad at high dose rate, much higher than the values of a few hundredths of a Mrad at the low
dose rates probed in the HE. However,
even though the geometry and construction of the 
scintillators studied in the in situ measurement and the measurements in Ref.~\cite{Biagtan1996125} are quite different,
and the dose rates quite different,
the functional dependence is similar.

\section{Conclusions}

We have looked at the dependence of the light loss for scintillator tiles installed
in the CMS endcap hadron calorimeter as a function of both dose and dose rate.  We see
a power law dependence consistent with that predicted by diffusion of oxygen (or other gas)
into the scintillator.  We see the functional form is in reasonable agreement with results
from gamma source irradiations both by us and by the authors of Ref.~\cite{Biagtan1996125}.

\acknowledgments

We would like to thank the CMS BRIL group for doing the FLUKA calculation and the staffs 
at the Michigan Memorial Phoenix source and the Argonne source for their help.
We congratulate our colleagues in the CERN accelerator departments for the excellent performance of the LHC and thank the technical and administrative staffs at CERN and at other CMS institutes for their contributions to the success of the CMS effort. In addition, we gratefully acknowledge the computing centres and personnel of the Worldwide LHC Computing Grid for delivering so effectively the computing infrastructure essential to our analyses. Finally, we acknowledge the enduring support for the construction and operation of the LHC and the CMS detector provided by the following funding agencies: BMWFW and FWF (Austria); FNRS and FWO (Belgium); CNPq, CAPES, FAPERJ, and FAPESP (Brazil); MES (Bulgaria); CERN; CAS, MoST, and NSFC (China); COLCIENCIAS (Colombia); MSES and CSF (Croatia); RPF (Cyprus); SENESCYT (Ecuador); MoER, ERC IUT and ERDF (Estonia); Academy of Finland, MEC, and HIP (Finland); CEA and CNRS/IN2P3 (France); BMBF, DFG, and HGF (Germany); GSRT (Greece); OTKA and NIH (Hungary); DAE and DST (India); IPM (Iran); SFI (Ireland); INFN (Italy); MSIP and NRF (Republic of Korea); LAS (Lithuania); MOE and UM (Malaysia); BUAP, CINVESTAV, CONACYT, LNS, SEP, and UASLP-FAI (Mexico); MBIE (New Zealand); PAEC (Pakistan); MSHE and NSC (Poland); FCT (Portugal); JINR (Dubna); MON, RosAtom, RAS and RFBR (Russia); MESTD (Serbia); SEIDI and CPAN (Spain); Swiss Funding Agencies (Switzerland); MST (Taipei); ThEPCenter, IPST, STAR and NSTDA (Thailand); TUBITAK and TAEK (Turkey); NASU and SFFR (Ukraine); STFC (United Kingdom); DOE and NSF (USA).


\bibliography{main}{}







\end{document}